\documentclass[12pt]{article}
\usepackage{amsmath,amstext,amsfonts,amsbsy,amssymb,amscd,bbm,epsfig}
\usepackage{graphicx}
\usepackage{subfigure}
\usepackage{courier}
\usepackage{fullpage}
=-0.5cm \oddsidemargin=-0.5cm

\newcommand{\be}{\begin{equation}}
\newcommand{\ee}{\end{equation}}

\begin{document}
\begin{titlepage}
\begin{flushright}
ROM2F/2006/18
\end{flushright}
\vskip2truecm

\begin{center}
\begin{large}
{\bf  The 4--D  Layer Phase as a Gauge Field Localization: Extensive Study of the
    5--D Anisotropic U(1) Gauge Model on the Lattice }
\vspace*{0.5cm}

\end{large}
P.~Dimopoulos$^{(a)}$\footnote{E-mail: dimopoulos@roma2.infn.it} 
K.~Farakos$^{(b)}$\footnote{E-mail: kfarakos@central.ntua.gr} and  
S.~Vrentzos$^{(b)}$\footnote{E-mail: vrentsps@central.ntua.gr} \vskip1truecm

%\today

%***************** {\it draft} *****************

\vskip1truecm
{\sl $^{(a)}$ INFN-Rome2 Universita di Roma 'Tor Vergata' \\
Dipartimento di Fisica I-00133, Rome, Italy\\}
{\sl $^{(b)}$ Physics Department, National Technical University\\
15780 Zografou Campus, Athens, Greece}
\end{center}
\vskip1truecm

\begin{abstract}
\noindent We study a $4+1$ dimensional pure Abelian Gauge model on the lattice with 
two anisotropic couplings independent of each other and of the coordinates. 
A first exploration of the phase diagram  using mean field approximation 
and monte carlo  techniques has demonstrated the existence of a  new phase, 
the so called Layer phase, in which the forces in the 4--D subspace are 
Coulomb--like while in the transverse direction (fifth dimension) the force
is confining. This allows the possibility of a gauge field localization 
scheme. In this work the use of bigger lattice volumes and higher statistics 
confirms the existence of the Layer phase and furthermore clarifies the issue 
of the phase transitions' order. We show that the Layer phase is 
separated from the strongly coupled phase by a weak first order phase 
transition. Also we provide  evidence that the Layer phase is separated 
by the five--dimensional Coulomb phase with a second order phase transition 
and we give a first estimation of the  critical exponents.

\end{abstract}
\end{titlepage}

\section{Introduction}
Higher dimensional theories have been introduced by Kaluza and Klein  
to achieve unification to all known, that time, interactions.
There is strong theoretical motivation for considering spacetimes with extra 
(more than three) spatial dimensions like String theory and M-theory that try to
incorporate quantum gravity in a consistent way. In the context of these theories 
ten (String theory) or eleven (M-theory)
spacetime dimensions are required.  
Although there is still a lack of experimental evidence on the existence
of a higher dimensional world, these  ideas
have shown a new revival during the last decade in the context
of brane world theories, the latter attempt to respond to 
long standing problems of theoretical physics like the hierarchy problem,
the cosmological constant problem and the fermion mass hierarchy.
Brane world theories assume that our world is a three brane
which is embedded in a higher dimensional space (bulk).

A class of these theories considers a (4+n)--dimensional space with n flat
compactified dimensions (ADD scenario)\cite{comp-extra} 
while a second class makes use of non-compact
warp extra dimensions (Randall-Sundrum first and second model) \cite{RS}.
 Although initially these theories were referring to the gravity interaction
they immediately gave rise to questions for the brane localization of the other
fields (for a review see \cite{rubakov},  \cite{csaki}, \cite{sundrum}).
 For the localization of fermions one can use the domain wall mechanism 
where a bounce-like static solution generated from some extra scalar 
field interact with the fermions.
Chiral fermions succeed to appear usually in that formulation 
\cite{rubakov}, \cite{fermions}. However there is a more powerfull mechanism 
where we can achieve localization 
of extended structures of particles which include gauge fields, fermions 
and scalar fields with gauge charge \cite{DS}, \cite{SU2} and is based on confinement 
along the extra dimensions. This mechanism may be triggered by the extra dimensional
gravity \cite{gravity}. 

Since the mid eighties  Fu and Nielsen proposed  a
five--dimensional abelian gauge lattice model  with anisotropic couplings that could  serve as
a new way of thought for achieving the  dimensional reduction \cite{funiel}. The idea was that
the anisotropy of the interactions between the four--dimensional space  and  the fifth (extra) dimension
could give a  phase diagram which contains a new kind of phase beyond the well known and
expectable strong and Coulomb phases. By using Mean Field methods it was shown that the new phase
was characterized as a Coulomb one in the four dimensions but confining along the remaining one.
This new phase was called Layer phase. Since the confinement along the extra dimension is responsible for 
the fact that there is no interaction between  neighbouring four--dimensional layers, that could serve as 
an indication of the effective existence of the  four dimensional world.

The higher dimensional gauge theories belong to the class of the non--renormalisable ones. Therefore 
such theories can be only valid  as effective  emerging from more fundamental theories the origin of which 
is still under discussion. One way to deal with these theories is to adopt the necessary existence 
of a cut-off $\Lambda$ and consider them as effective theories  for low enough energies. 
In any case the perturbation theory seems not to be sufficient to describe a mechanism for 
the gauge field localization on three dimensional submanifold for which the interaction of the 
gauge and matter fields along the extra dimensions must be suppressed. 
With  the five--dimensional anisotropic gauge 
abelian model that we study here we intend to present   a gauge field localization mechanism 
on the lattice realised by means of the Layer phase.  The property of confinement along the extra dimension 
which characterizes the Layer phase has to be studied using necessarily  non--perturbative tools.

A first numerical study of the model has been done in \cite{stam}  in which the Layer phase 
was identified by means of Monte--carlo techniques. In \cite{DFKK} the same model was studied in 
more extent and furthermore a new version was presented consisting of defining the coupling anisotropy as 
dependent of the extra dimension coordinate. This version of the anisotropic model was inspired by  the mechanism 
used to establish the higher dimensional models with warp extra dimensions mentioned above \cite{RS}. 
It also provided the possibility of having the Layer phase on the lattice. 

The aim of this paper is to study more intensively  the phase diagram of the model in terms of  two gauge couplings,
one defined on the four--dimensional subspace and the other along the extra (transverse) direction. For that we 
employ bigger lattice volumes and  higher statistics than  used in the past. 
Our purpose is to show that the Layer phase not only can be identified with  precision using 
the lattice techniques but furthermore to demonstrate that it is well separated both from
the five--dimensional Coulomb phase and the confining phase of the model.  Actually we bring results that
are in favour of a first order phase transition  between the Layer and the strong phase.
Moreover  we verify that this conclusion is also valid in the limit of the very strong couplings along 
the extra fifth dimension for which   the features of the  strong--Coulomb transition for the
four--dimensional abelian gauge model are reproduced  (see Section 4.1). 
On the other hand we provide strong evidence that the 
Layer--Coulomb (5D) transition is of second order (see Section 4.2). 
Although the lattice volumes that we have been able to 
use  appear not to be  sufficient to give a definite and conclusive  answer on the problem of the order of the 
phase transition, 
nevertheless, we are allowed to extract a first estimation of the critical exponents.

\section{The model}
We consider the  U(1) gauge lattice action in five dimensions with two anisotropic
couplings $\beta$ and $\beta^{'}$:

\begin{equation}\label{action}
S^{5D}_{gauge}=\beta \sum_{x,1 \leq \mu < \nu \leq 4}(1- Re(U_{\mu\nu}(x))) +
\beta^{'} \sum_{x,1 \leq \mu \leq 4}(1- Re(U_{\mu T}(x)))
\end{equation}
where
\begin{eqnarray}
U_{\mu\nu}(x)&=&U_{\mu}(x)~U_{\nu}(x+a_{s}\hat{\mu})~U_{\mu}^{\dagger}(x+a_{s}\hat{\nu})~U_{\nu}^{\dagger}(x) \nonumber \\
U_{\mu T}(x)&=&U_{\mu}(x)~U_{T}(x+a_{s}\hat{\mu})~U_{\mu}^{\dagger}(x+a_{T}\hat{T})~U_{T}^{\dagger}(x) \nonumber
\end{eqnarray}
are the plaquettes defined on the 4--D subspace ($\mu,\nu=1,2,3,4$) and on the plane containing the transverse 
direction ($x_{T}$) respectively. We also denote with $a_{S}$ and $a_{T}$ the corresponding lattice 
spacings. The link variables are given by $U_{\mu}=e^{i \theta_{\mu}}$  for the 
4--D subspace and $U_{T}=e^{i \theta_{T}}$ for the transverse direction \footnote{In terms of the continuum fields 
they would be written as $\theta_{\mu}(x) = a_S A_{\mu}(x)$ and $\theta_{T}(x) = a_T A_{T}(x)$.}.
The plaquettes can also be written in the following form:
$$ U_{\mu\nu}(x) = \exp(i \theta_{\mu \nu}(x)), ~~~~ U_{\mu T}(x) = \exp(i \theta_{\mu T}(x)) $$
with the definitions : \\
\begin{eqnarray}
\theta_{\mu \nu} &=&  \theta_\mu(x) + \theta_\nu(x + a_{S} \hat{\mu}) - \theta_\mu(x+a_S \hat{\nu}) - 
\theta_\nu(x) \nonumber \\
\theta_{\mu T} &=& \theta_\mu(x) + \theta_{T}(x + a_{S} \hat{\mu}) - \theta_\mu(x+a_T \hat{T}) - \theta_T(x) \nonumber
\end{eqnarray}

\subsection{Observables}
Two operators, which are  mainly used to define the different phases of the model
and  help to estimate the order of the phase transitions, are the space--like, $\hat{P_{S}}$,
 and the time--like plaquette, $\hat{P_{T}}$ and  are given by the following 
expressions:

\begin{equation}\label{Ps}
 \hat{P_{S}}\equiv\frac{1}{6L^{5}}\sum_{x,1\leq \mu < \nu \leq 4} \cos \theta_{\mu \nu}(x)
\end{equation}

\begin{equation}\label{Pt}
\hspace*{-0.5cm} \hat{P_{T}}\equiv\frac{1}{4L^{5}}\sum_{x,1\leq \mu \leq 4} \cos \theta_{\mu T}(x)
\end{equation} 
where  $L$  is the length of each lattice direction which  is assumed to be the same in all directions.

\noindent Starting from  the operators (\ref{Ps}) and (\ref{Pt}) we measure the following quantities:

\begin{enumerate}
\item The  space-like plaquette mean value:
\begin{equation}
P_{S}\equiv <\hat{P_{S}}>
\end{equation}
where the symbol $ < \ldots > $ denotes the statistical average with action given by Eq.(\ref{action}).
\item The transverse-like plaquette mean value:
\begin{equation}
P_{T}\equiv <\hat{P_{T}}>
\end{equation}

\item The distributions $N(\hat{P_{S}})$, $N(\hat{P_{T}})$ of $\hat{P_{S}}$ and $\hat{P_{T}}$ respectively.
\item The susceptibilities of $\hat{P_{S}}$ and $\hat{P_{T}}$ :
\begin{equation}
S(\hat{P_{S}})=V(< \hat{P_{S}^{2}} > - < \hat{P_{S}}>^{2}),~~~~ 
S(\hat{P_{T}})=V(< \hat{P_{T}^{2}} > - < \hat{P_{T}}>^{2})
\end{equation}
where $V$ stands for the  lattice volume in five or four dimensions  depending on the case under study
(see details below).
\item The Binder cumulants of $\hat{P_{S}}$ , $\hat{P_{T}}$ :
\begin{equation}
B(\hat{P_{S}})\equiv 1-\frac{<\hat{P_{S}}^{4} >}{3 < \hat{P_{S}}^{2} >},~~~~
 B(\hat{P_{T}})\equiv 1-\frac{<\hat{P_{T}}^{4} >}{3 < \hat{P_{T}}^{2} >}
\end{equation}
\end{enumerate}

Furthermore we use the  helicity modulus (h.m.) first  introduced in the context of 
lattice gauge theories in \cite{vetfor}. It  is an order parameter which characterizes 
the response of a system to an external electromagnetic flux. 
More precisely  it is the curvature of the flux free energy (F($\Phi$)) at the origin 
(in fact any point $\Phi_{0}$ different from the origin will work  fine):
\begin{equation}\label{def_hs}
h(\beta)=\left.\frac{\partial^{2}F(\Phi)}{\partial\Phi^{2}}\right|_{\Phi=0}
\end{equation}
The h.m. takes  always zero value  in the confined phase and values different from zero in 
the Coulomb phase.
Since in our model the Layer phase  is a mixture of both a confining and Coulomb phase,
we propose the following measuring procedure:\\
we impose first the extra flux $\Phi$ on a stack of plaquettes 
(following \cite{vetfor})
\begin{equation}
stack= \lbrace\theta_{\mu \nu} | \mu=1,\nu=2;x=1,y=1\rbrace
\label{stack}\end{equation}
\noindent then with a change of variables we spread the extra flux,
uniformly, to all the plaquettes with the given orientation. Now, our
partition function  becomes:
\begin{equation}\label{Zfun}
 Z(\Phi)=\int D\theta~~{\large e^{\beta\sum_{(\mu \nu)-planes} 
\cos(\theta_{\mu \nu}+
\frac{\Phi}{L_{\mu}L_{\nu}}) + \beta \sum_{(\overline{\mu \nu})-planes} 
\cos(\theta_{\mu \nu})} ~ 
e^{\beta^{'}\sum_{1\leq \mu \leq 4}\cos(\theta_{\mu T})}}
\end{equation} 
where $\sum_{(\mu \nu)}$  denotes the sum over all planes parallel to a
given orientation and $\sum_{(\overline{\mu \nu})}$  stands for the sum over the remaining planes.\\
The flux free energy is defined by
\begin{equation}
F(\Phi)=-\log\frac{Z(\Phi)}{Z(0)}
\end{equation}
In this way, using the h.m. definition (\ref{def_hs}), we obtain:
\begin{equation}
h_{S}(\beta)=\frac{1}{(L_{\mu}L_{\nu})^{2}} \left ( \left <\sum_{P}(\beta~ \cos(\theta_{\mu \nu}))\right > -
\left <(\sum_{P}(\beta~ \sin(\theta_{\mu \nu})))^{2} \right >\right )
\end{equation}
where the sum extends to all the plaquettes in the $(\mu \nu)$ orientation and the 
brakets denote the average over the gauge ensemble according to the partition function (\ref{Zfun}) with $\Phi=0$.

In a similar way  if we choose the ($\mu$,T) orientation and follow exactly the same 
steps described above for the space--like h.m., we obtain the expression for the ``transverse'' h.m.:
\begin{equation}
h_{T}(\beta^{'})=\frac{1}{(L_{\mu}L_{T})^{2}} \left ( \left < \sum_{P^{'}}(\beta^{'} \cos(\theta_{\mu T}))\right >
-\left <(\sum_{P^{'}}(\beta^{'} \sin(\theta_{\mu T})))^{2} \right > \right )
\end{equation}
The  sum now  extends to all the plaquettes on the transverse plane.

\section{The Phase diagram}
Before proceeding to the detailed analysis, we present in advance  the 
model phase diagram and a general description of the behaviour of the 
quantities that are used to specify the features of the phases \footnote{ 
For the mean field prediction  of the phase diagram 
see \cite{funiel}. Also for previous attempts for the phase diagram prediction  using 
numerical simulations see \cite{stam}, \cite{DFKK}.}. The phase diagram is
depicted in Fig.(\ref{Ph_diag}). Full triangles   represent 
the results obtained with  hysteresis loop 
study on $6^5$ and $8^5$ lattice volumes regarding the space--like and the 
transverse--like plaquette. For the points shown with   ``squares'' instead an
extensive high statistics analysis has been performed.
The phase diagram includes three distinct phases. 
For large values of $\beta$ and $\beta^{'}$ the model lies in a Coulomb phase 
($\mathbf{C}$) on the 5--D space. 
Now, if $\beta$ is kept constant, above the value of one, 
while  $\beta^{'}$ decreases the system  will eventually
show up a behaviour according to which  the force in  four dimensions  
will still be Coulomb-like 
while in the fifth direction the property of confinement is present.
This is the new phase   called Layer phase ($\mathbf{L}$). 
For small values of both $\beta$ and $\beta^{'}$, the force will be
confining in all five directions and the corresponding phase is the  Strong phase 
($\mathbf{S}$).
According to this way of reasoning two test charges found in the Layer phase will experience 
a Coulomb force in four dimensions with coupling given by the four--dimensional coupling
$\beta$,  while along the fifth direction they will experience a strong force as the corresponding
coupling $\beta^{'}$ takes small values. Therefore the Layer phase  can provide us with a mechanism 
for gauge field localization on a 4--D subspace in the context of higher dimensional models.
Since the potential between two charges can be expressed by the Wilson loops we will expect 
the following behaviour \cite{funiel}:

\begin{figure}[!h]
\begin{center}
{\includegraphics[width=12cm,angle=270]{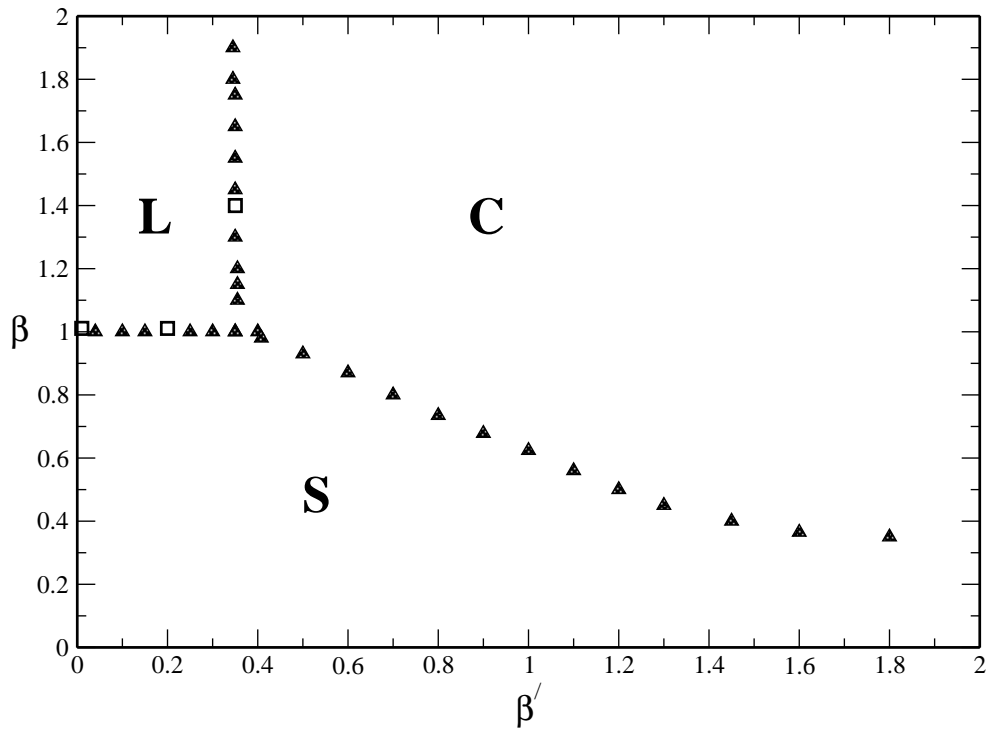}}
\caption{The phase diagram of the 5--D anisotropic  U(1) gauge model.}
%\label{phase_diag}
\label{Ph_diag}
\end{center}
\end{figure}  

\begin{itemize}
\item $W_{\mu \nu}(L_{1},L_{2}) \simeq \exp(-\sigma L_{1}L_{2})$ ~~(Confinement phase, $1\leq \mu,\nu \leq 5$))
\item $W_{\mu \nu}(L_{1},L_{2}) \simeq \exp(-\tau(L_{1}+L_{2}))$ ~~(Coulomb phase, $1\leq \mu,\nu \leq 5$)
\item $W_{\mu \nu}(L_{1},L_{2}) \simeq \exp(-\tau^{'}(L_{1}+L_{2}))$  ~~ and
\item $W_{\mu T}(L_{1},L{2}) \simeq exp(-\sigma^{'}L_{1}L_{2})$ ~~~~~~
(Layer phase,$1 \leq \mu,\nu \leq 4$.)
\end{itemize}

\noindent Moreover if we consider  the helicity modulus we find that it shows  the following properties:\\
(i) In the   Strong  phase (keeping $\beta^{'}$ constant) the space--like helicity modulus,
$h_{S}(\beta)$,  takes zero value  and as we approach and eventually pass the
critical point it must become non-zero in the Layer phase with a value that approaches 
one as $\beta$ increases futher.
On the other hand the transverse--like h.m., $h_{T}(\beta^{'})$, must remain zero for all values 
of $\beta$ since both phases exhibit confinement in the fifth direction (see Section 4.1).\\
(ii) For the transition between the Coulomb and the Layer phase we expect for
$h_{S}(\beta)$ to get a value close to one for all the values of
$\beta^{'}$ since the 4-dimensional layers are in a Coulomb phase,
while $h_{T}(\beta^{'})$ gets  zero value for the
Layer phase and as we pass the critical point and enter the Coulomb
phase it must grow towards one as  $\beta^{'}$ increases (see Section 4.2).

\section{Monte carlo Results}
We used a 5-hit Metropolis algorithm supplemented by an overrelaxation method (see ref.\cite{DFKK}
and references within). The lattice volumes used were: $6^5$, $8^5$, $10^5$, $12^5$  and 
$14^5$. More than $10^5$ sweeps were dedicated to the thermalisation process and we got 
samples of about $5-9 \times  10^4$ measurements free of autocorrelation. 
Also two self-adjusting scales were implemented, 
one for the update procedure on the 4--D subspace  and the other along the
transverse dimension. 
The errors of the various measured quantities  have been calculated with the 
jackknife method.

In the following sections we will study the Strong--Layer and the Coulomb--Layer phase transitions which are of 
main interest. In this work we will not study the   
Strong--Coulomb phase transition. However we note that strong evidence for a  first order phase transition has 
been found due to pronounced two peak distributions (see \cite{DFKK}).  

\subsection{Strong-Layer Phase Transition}
\begin{figure}[!h]
\begin{center}
\subfigure[]{\includegraphics[scale=0.30,angle=-90]{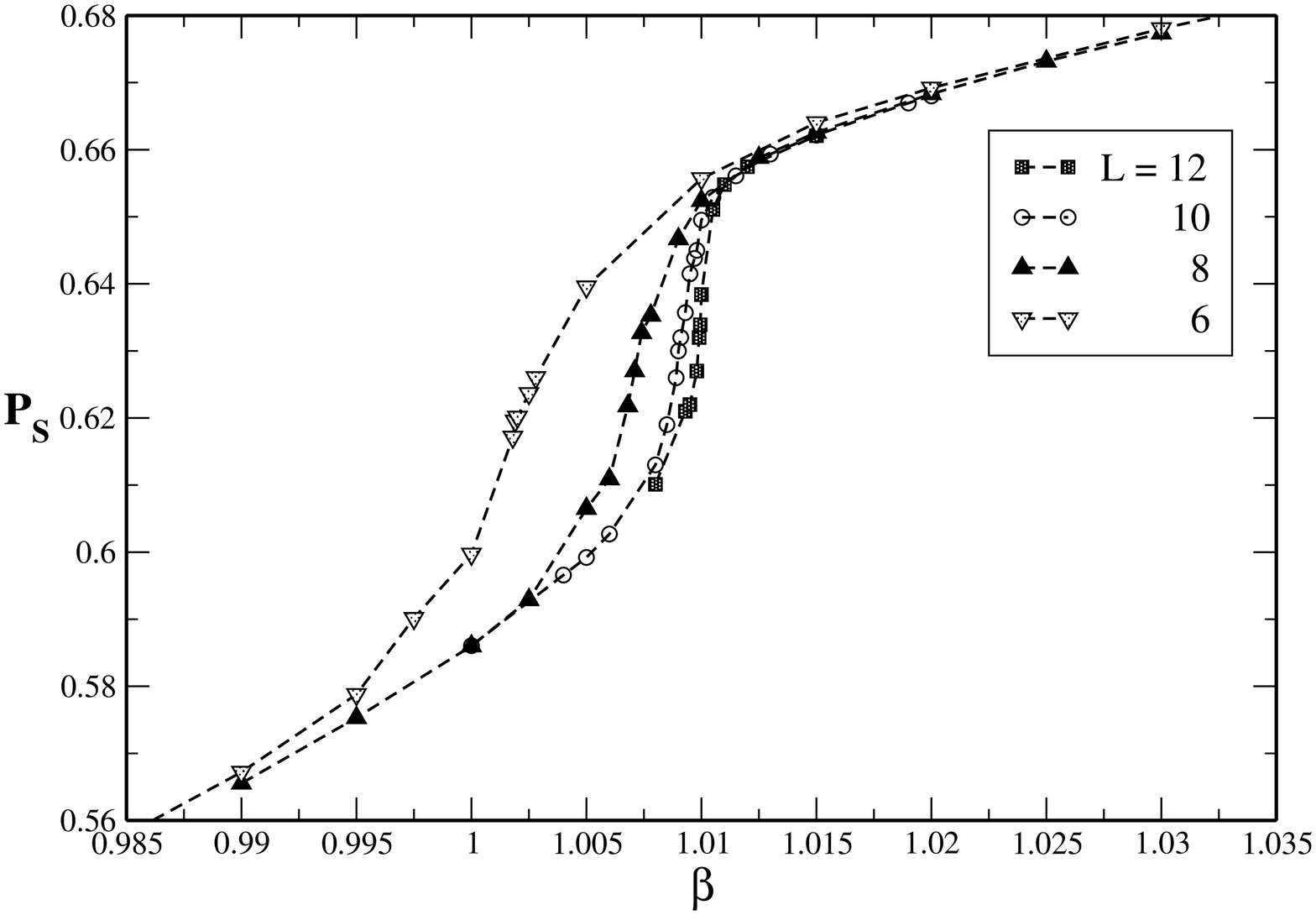}}
\subfigure[]{\includegraphics[scale=0.30,angle=-90]{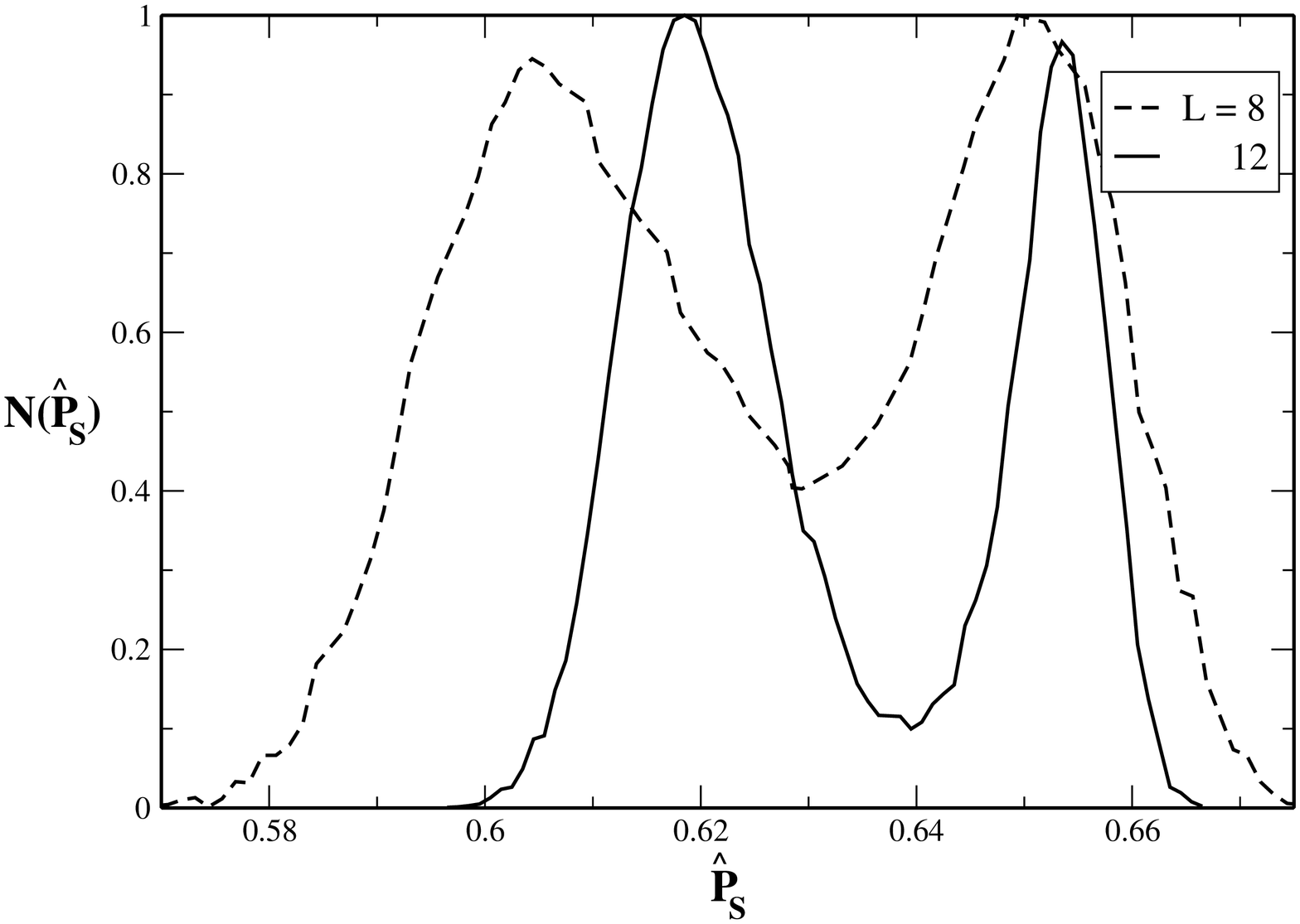}}
\caption[]{(a): Mean values for the space-like plaquette 
(the errors are included in the symbols' size and the dashed lines guide the trend); 
(b): The two peak distribution 
$N(P_{S})$ for L=8 at  $\beta^{'}=0.2$ and $\beta=1.0073$ and L=12 at $\beta^{'}=0.2$ and
$\beta=1.0099$.}
\label{PS_distr}
\end{center}
\end{figure}

We choose a constant value for the  coupling $\beta^{'}= 0.2 $
and we let the value of $\beta$ vary for four lattice volumes $6^5, 8^{5},10^{5} ~\mbox{and}~ 12^{5}$. 
For low enough values of $\beta$ the $P_{S}$ tends to values  equal to $\beta / 2$ according to the 
strong coupling expansion, then grows as it passes to the Coulomb phase  tending to values equal to 
the weak coupling limit $1 - 1/(d~\beta)$ (see Fig.\ref{PS_distr}). 
The transition becomes  steeper as the lattice 
volume increases \footnote{Note that $P_{T}$ remains constant to the strong coupling
value i.e. $\beta^{'}/2=0.1$ during the transition (see also \cite{DFKK}).}. 
A first  evidence for a first order phase transition can be found in Fig.\ref{PS_distr} 
where is shown a two state signal for the space--like plaquette which persists and  becomes more 
pronounced as we pass from lattice volume $8^5$ to $12^5$. We should note that the two--state 
signal is present only when  we study the gauge invariant quantities measured on the 
four dimension layers and not on the whole volume. The reason is that as the system passes to the 
Layer phase with a non-continuous way the various quantities measured on the layers show  a
"non-coherent" behaviour. This phenomenon, in the case of a strong 
first order phase transition, is responsible for producing multipeak distributions 
for the quantities measured on the whole five--dimensional lattice volume 
while for a weak first order phase transition  one peak wide distributions are formed. 
Based on that observation and in order to obtain a more explicit  signal  we  study the phase transition 
on the layer 
\footnote{The same has been found and identified in the case of the Layer--Higgs 
phase \cite{DF}.}. 
  
\begin{figure}[!h]
\begin{center}
\includegraphics[angle=270,width=10cm]{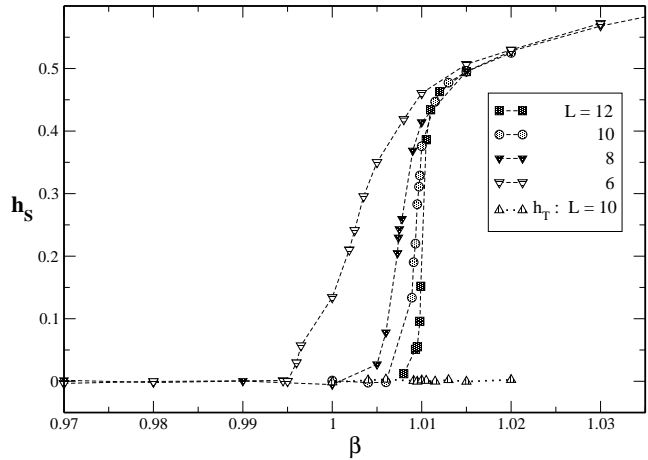}
\caption{The volume dependence of  the space--like h.m. $h_{S}$. The 
transverse h.m. $h_{T}$ (uptriangles) for L=10 is also shown 
(the errors are included in the symbols' size and the dashed lines guide the trend).}
\label{hs_fig}
\end{center}
\end{figure}

The behaviour of the space--like helicity modulus, $h_{S}$,  for the same transition is depicted   
in Fig.\ref{hs_fig}. As it was expected the $h_S$ takes values strictly around zero in the 
confinement phase and passes to non-zero values in the Coulomb phase. The transition 
shows a steeper passage as the lattice volume takes bigger values. In particular what is to be noticed  
is that for the bigger lattice volume a rather high jump arises around $\beta \sim 1.01$. 
In the same figure we have included the $h_T$  for one volume which takes zero value 
in both phases (shown with ``uptriangles'') which is an imprint  of the confinement along the 
fifth direction.

\begin{figure}[!h]
\begin{center}
\subfigure[]{\includegraphics[scale=0.30,angle=-90]{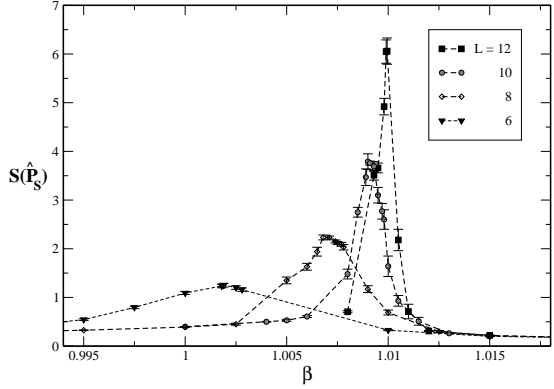}}
\subfigure[]{\includegraphics[scale=0.30,angle=-90]{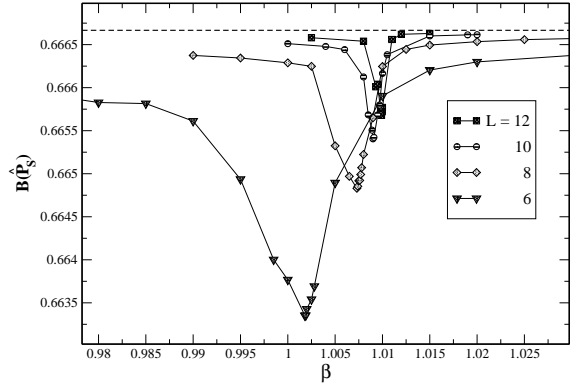}}
\caption{ Volume dependence of the susceptibility (a) and of the 
Binder cumulant (b) for the space-like plaquette (the lines guide the trend).} 
\label{susc_cum}
\end{center}
\end{figure}

The volume dependence of the susceptibility $S(\hat{P_S})$  and of the Binder cumulant, 
$B(\hat{P_s})$, are 
illustrated in Fig.\ref{susc_cum}. 
The  $S(\hat{P_s})$ (measured on the 4--D subspace according to what has been noted above) 
exhibits a clear  increase with the volume but it is not a linear one. The minimum of the Binder cumulant
also tends to increase  with the volume though  slowly. For the bigger lattice volume used (i.e. $12^5$) 
the minimum value, $B^{min}(\hat{P_s})$,  seems to lie rather 
far from the 2/3 which should be the infinite volume limit in the case   
of a higher order phase transition \cite{Binder}. 

\begin{figure}[!t]
\begin{center}
\includegraphics[angle=270,width=10cm]{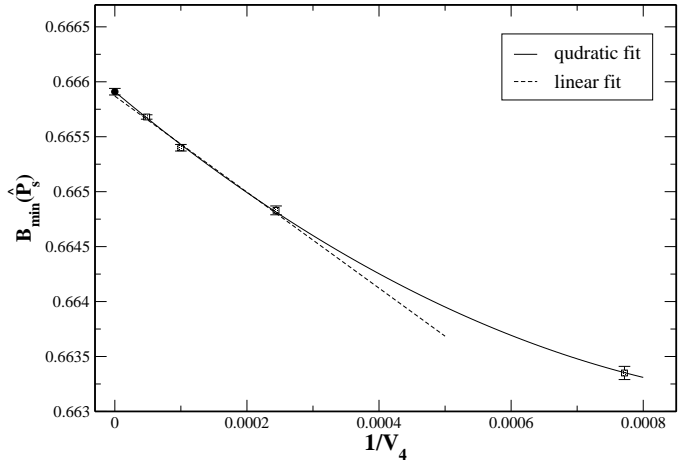}
\caption[]{Fit for $B_{min}(\hat{P_{S}})$ to the infinite volume.}
\label{inf_cum}
\end{center}
\end{figure}

We attempt to estimate the infinite volume limit for the minimum of the Binder cumulant, 
$B_{min}(\hat{P_s})$, and for that we use the ansatz:
$$
B_{min}(\hat{P_{S}})=B_{min, \infty}+\sum_{k=1}^{\infty}\frac{C_{k}}{V_4^{k}}
$$
where $V_4$ is for the layer volume and  we limit ourselves to $k=2$. 
The relevant values for the fit are given in Table \ref{beta_prime_02}. In Fig.\ref{inf_cum} we depict the 
fit. The infinite volume result we obtain is:
$ B_{min, \infty} = 0.66591(3)$ which is well far from 2/3.   
Note in passing that a linear fit of the three higher values  gives compatible result. So  we come to 
the conclusion that a  higher order phase transition seems not to be the case.

Now we try to apply a finite size analysis for the maxima of the susceptibility 
of the space--like plaquette.
Following  the method proposed in \cite{BK} and also applied in the 4--D U(1) gauge model in 
\cite{ArLiSc}, we set the ansatz:
\begin{eqnarray}\label{fit_susc}
\beta_{c} &=& \beta_{\infty}+\sum_{k=1}^{\infty}\frac{A_{k}}{V_4^{k}} \nonumber \\
\frac{\beta_c^{2} S_{max}(\hat{P}_{s})}{V_4} &=& \frac{1}{4}G_{\infty}^{2}+\sum_{k=1}^{\infty}\frac{B_{k}}{V_4^{k}}
\end{eqnarray}

\noindent where:  $\beta_{c}$ stands for the pseudocritical value of the coupling  $\beta$ at each lattice volume,
$\beta_{\infty}$ is the coupling value at infinite volume and $G_{\infty}$ is the infinite volume gap 
in the plaquette energy.
The $\beta_{c}$ values have been estimated by making a gaussian fit around the peak of the susceptibility
while the use of the multihistogram method provided compatible results.
The values used for the fits  are given in the Table \ref{beta_prime_02}. 

\begin{table}
\begin{center}
\begin{tabular}{cccccc}
\hline \hline 
 $L$ & $\beta_c$ & $B_{min}(\hat{P_{S}})$ &  $S_{max}(\hat{P}_{s})$ &  $E_{S}$ & $E_{L}$\\
\hline
$6$  & 1.00180(6) & 0.66335(6)  & 1.125( 8) & 0.5925(7) & 0.6484(8)\\
$8$  & 1.00710(4) & 0.66483(4) & 2.254(50) & 0.6065(2) & 0.6511(2) \\
$10$ & 1.00910(9) & 0.66540(3) & 3.784(40) & 0.6141(2) & 0.6528(1)\\
$12$ & 1.00995(3) & 0.66568(2) & 6.110(30) & 0.6190(1) & 0.6535(1) \\
\hline
\end{tabular} \vspace*{0.3cm}
\caption{\label{beta_prime_02} The pseudocritical gauge coupling values, the Binder cumulant minima, 
$B_{min}(\hat{P_{S}})$, the maxima of the space--like 
susceptibility,  $S_{max}(\hat{P}_{s})$,  and  the energy peaks 
in the Strong and Layer phase for $\beta^{'}=0.2$. }
\end{center}
\end{table}

The results obtained from  the fitting procedure with  $k=2$ in Eq.(\ref{fit_susc})   are: 
\begin{equation}\label{beta_Ginfty}
\beta_{\infty}=1.01072(5) ~~~\mbox{and}~~~ G_{\infty}=0.0297(5) 
\end{equation}
The value for $\beta_{\infty}$ lies close to the critical value 
of the coupling in the 4--D U(1) gauge model found in \cite{ArLiSc}. This is rather logical
since in both models there is a transition from a confinement phase to a 4-D Coulomb phase. 
This fact is also consistent with the assumption that the 4--D layers pass to the Coulomb phase 
in a more or less independent way since the transverse interaction is of a strong type.

\begin{figure}[!h]
\begin{center}
\subfigure[]{\includegraphics[scale=0.30,angle=-90]{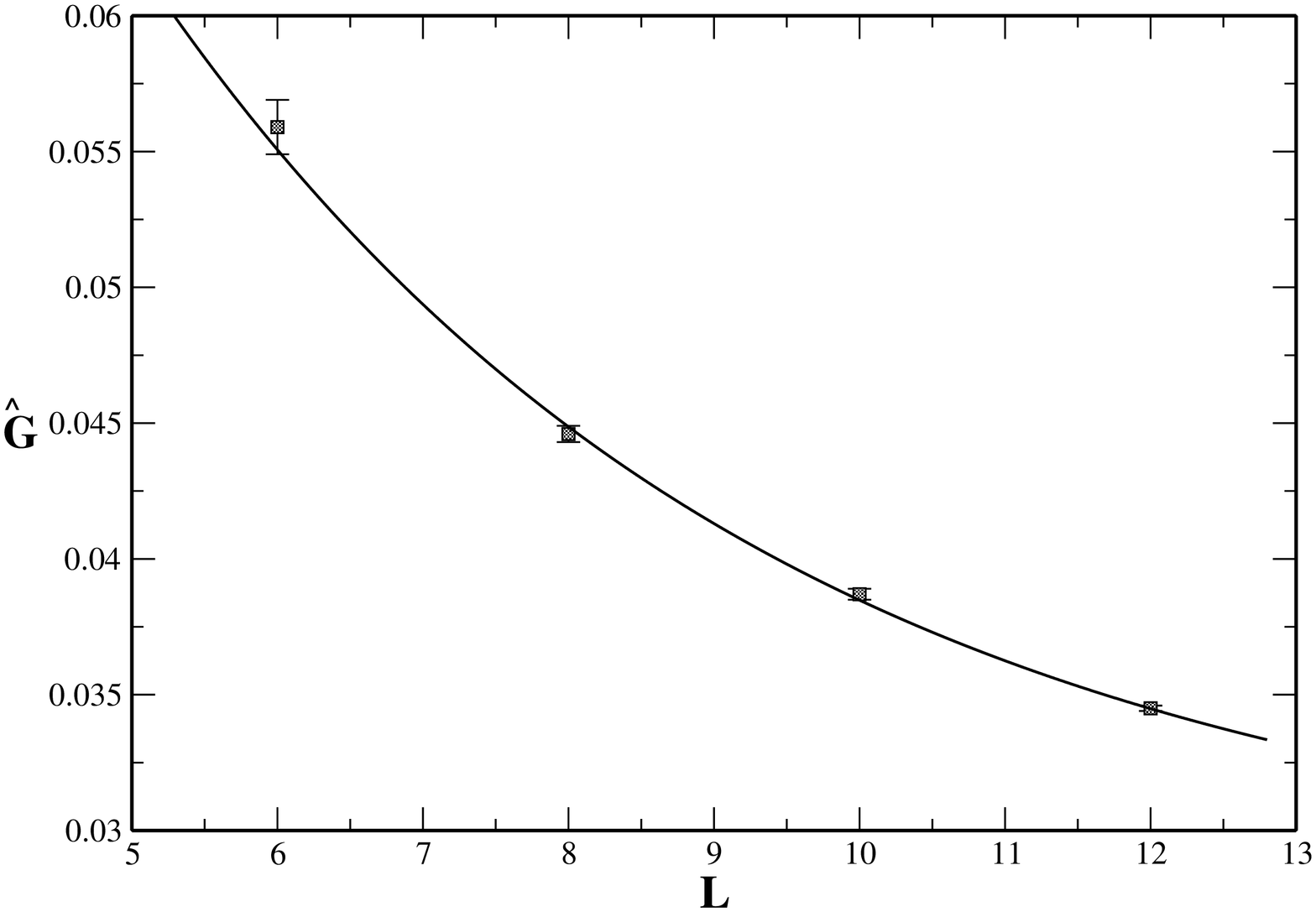}}
\subfigure[]{\includegraphics[scale=0.30,angle=-90]{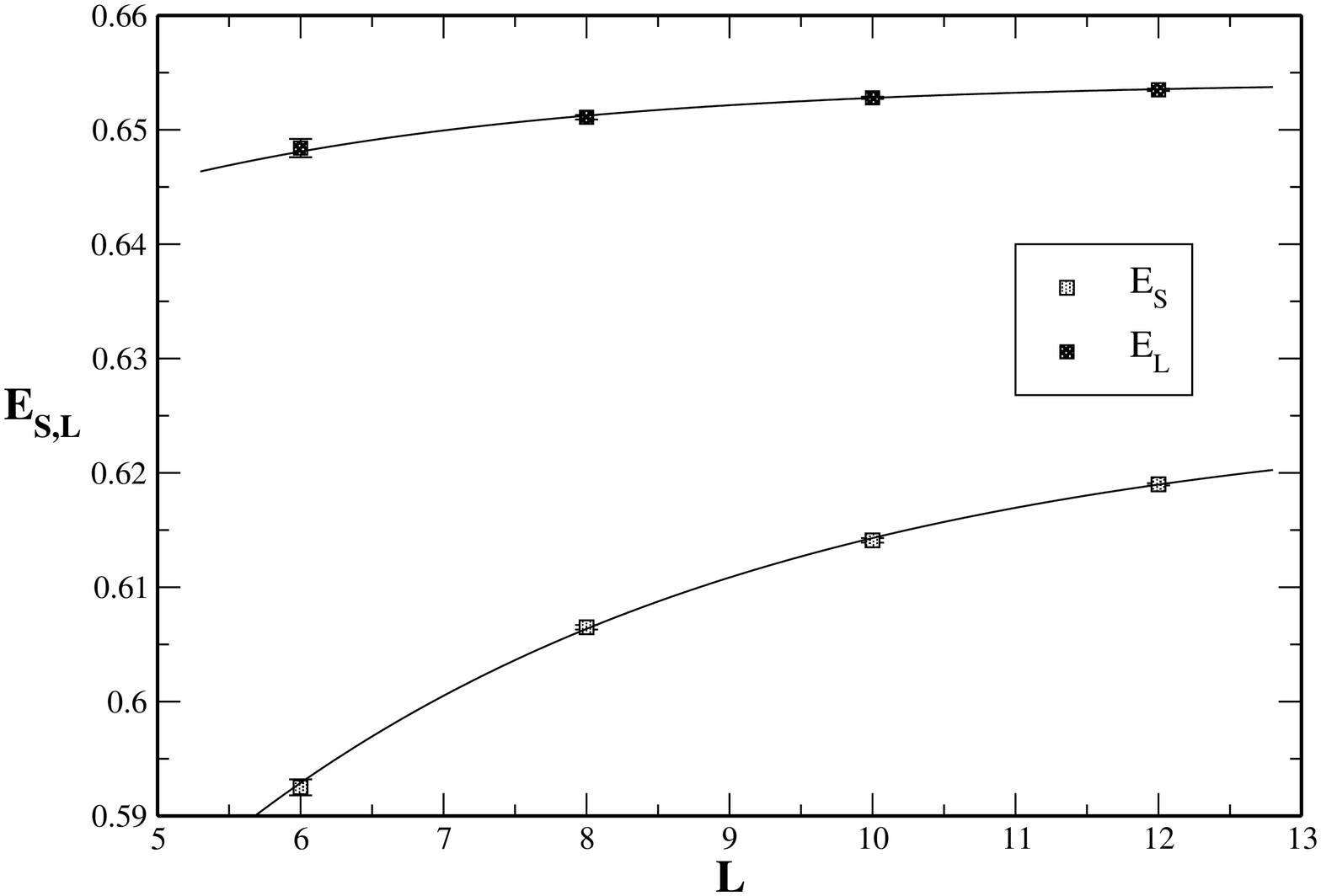}}
\caption[]{(a): The fit for the energy Gap, $\hat{G}$, according to the Eq.(\ref{gap_fit}); 
(b): The fits for the energy peaks according to the Eq.(\ref{E_LS}).}
\label{G_hat}
\end{center}
\end{figure}

The infinite volume gap can be measured in two additional ways. 
First from the double peak histograms we estimate  the values  of the energy peaks $E_{i=S,L}$
which correspond to the Strong and Layer phases, using independent Gaussian fits in the vicinity of 
each peak. Then we calculate the difference $\hat{G} = E_L - E_S$  for each lattice volume and we 
make use of the ansatz:
  
\begin{equation}\label{gap_fit}
\hat{G}=G_{\infty}+G_{1} e^{-G_{2} L}
\end{equation}
The fit is depicted in Fig.\ref{G_hat}(a). In this way we  obtain  the value for the infinite volume 
gap,  

\begin{equation}\label{G2}
G_{\infty}=0.0278(13)
\end{equation}

Alternatively by using the energy peaks  $E_{i=S,L}$ found above and given in Table \ref{beta_prime_02} 
we perform exponential fits of the form:
\begin{equation}\label{E_LS}
E_{i}(L) = E_{i, \infty} + c_{1,i}e^{-c_{2,i} L} ~~\mbox{where}~~ i=S,~L 
\end{equation}
The results are: $E_{S \infty} = 0.6257(8)$ and $E_{L \infty} = 0.6543(4)$ from which we obtain the 
energy gap equal to: 
\begin{equation}\label{G3}
G_{\infty} = 0.0286(9)
\end{equation}
 
The comparison between Eqs.(\ref{beta_Ginfty}), (\ref{G2}) and (\ref{G3}) allows the conclusion  that the 
three different methods applied for the energy gap calculation give compatible 
results and safely far from zero.   

It would be interesting to move to a different value 
of the transverse coupling in order to have a broader view of the features of the phase transition. 
We choose a rather small value, namely, $\beta^{'}=0.01$. This choice is justified by the fact that  
it brings us  closer to the 4--D case for which a weak first order phase transition has been found
\cite{ArLiSc}. We repeat the same analysis as for the $\beta^{'}=0.2$ case. The relevant values for the
pseudocritical gauge couplings, the maxima of the susceptibility, the energy gap and  the 
energy peaks $E_{S,L}$  are given in Table \ref{beta_prime_001}.  

\begin{table}
\begin{center}
\begin{tabular}{ccccc}
\hline \hline 
 $L$ & $\beta_c$ &  $S_{max}(P_{S})$ &  $E_{S}$ & $E_{L}$ \\
\hline
$6$  & 1.00190(8) & 1.125(20) & 0.5900(9)& 0.6467(8)  \\
$8$  & 1.00750(6) & 2.210( 3) & 0.6061(7)& 0.6510(6) \\
$10$ & 1.00930(4) & 3.710( 9) & 0.6144(5)& 0.6531(4) \\
$12$ & 1.01010(3) & 5.980(34) & 0.6189(3)& 0.6336(3) \\
\hline
\end{tabular} \vspace*{0.3cm}
\caption{\label{beta_prime_001} The pseudocritical gauge coupling values, the maxima of the space--like 
susceptibility,  $S_{max}(P_{S})$,    and  the values of the energy peaks 
in the Strong and Layer phase for $\beta^{'}=0.01$. }
\end{center}
\end{table}   

\begin{table}
\begin{center}
\begin{tabular}{ccc}
\hline \hline 
 $\beta^{'}$ & $\beta_{\infty}$ &  $G_{\infty}$ \\
\hline
0.2 & 1.01072(5) & 0.0297(5)  / 0.0278(13) / 0.0286(9) \\
0.01& 1.01077(5) & 0.0305(20) / 0.0294(23) / 0.0303(14) \\
\hline
\end{tabular} \vspace*{0.3cm}
\caption{\label{all} The critical values for the space--like gauge coupling 
and the values for the infinite volume energy gap 
calculated in three ways as explained in the text, for two values of the 
transverse gauge coupling namely, $\beta^{'}=0.2, 0.01$.}  
\end{center}
\end{table}   

Our results for both $\beta^{'}=0.01$ and $\beta^{'}=0.2$ can be found in Table \ref{all} and lead to 
two conclusions. The first is that the phase transition occurs at almost constant value for the 
spatial gauge coupling $\beta$. Moreover the critical value for the transition 
between the 5--D confinement phase to the Layer phase lies very close to the corresponding critical value 
found for the 4--D U(1) gauge model. The second conclusion is that due to the non zero value for the 
infinite volume energy gap combined 
with all the rest of the analysis done, we have a  clear evidence  for a first order phase  transition 
though a weak one.

\subsection{Coulomb-Layer Phase Transition}
We used four lattice volumes \footnote{The use of  rather small lattice volumes such as $4^5$ and $6^5$
gives non reliable information in the light of the higher volume 
results. This fact is mainly responsible for extracting the conclusion of a possible
crossover in \cite{DFKK}.}, namely: $8^{5}$,$10^{5}$,$12^{5}$ and $14^{5}$.
The gauge invariant quantity used for this transition is the transverse--like plaquette 
whose values in the confinement phase tend to the strong coupling limit, $\beta^{'}/2$, 
and grow as the system passes to the Coulomb phase. The space--like plaquette, as the forces on the
4--D subspace are of Coulomb type does not show any substantial change of its value (see \cite{DFKK}).
 
\begin{figure}[!h]
\begin{center}
\subfigure[]{\includegraphics[scale=0.30,angle=-90]{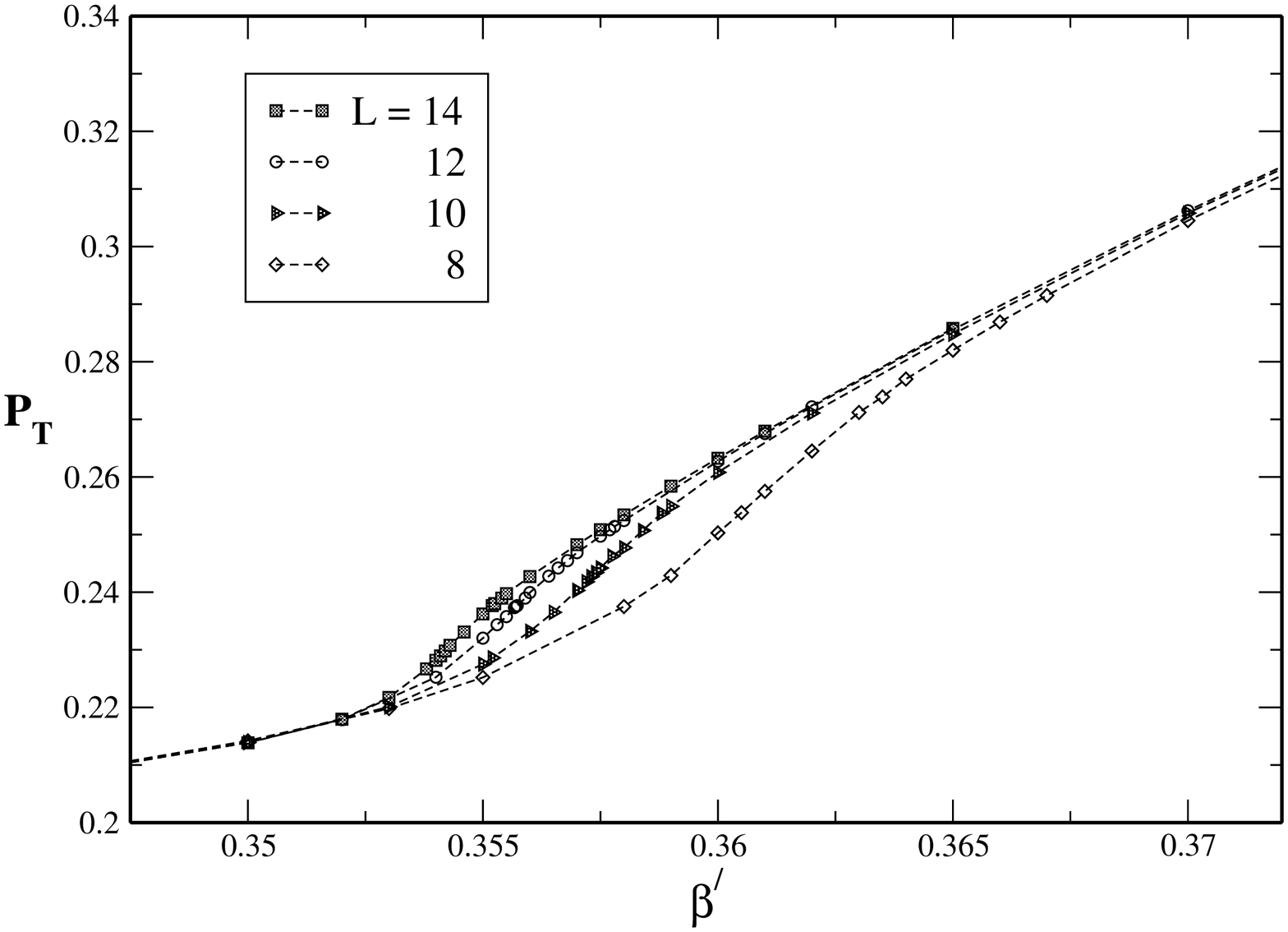}}
\subfigure[]{\includegraphics[scale=0.30,angle=-90]{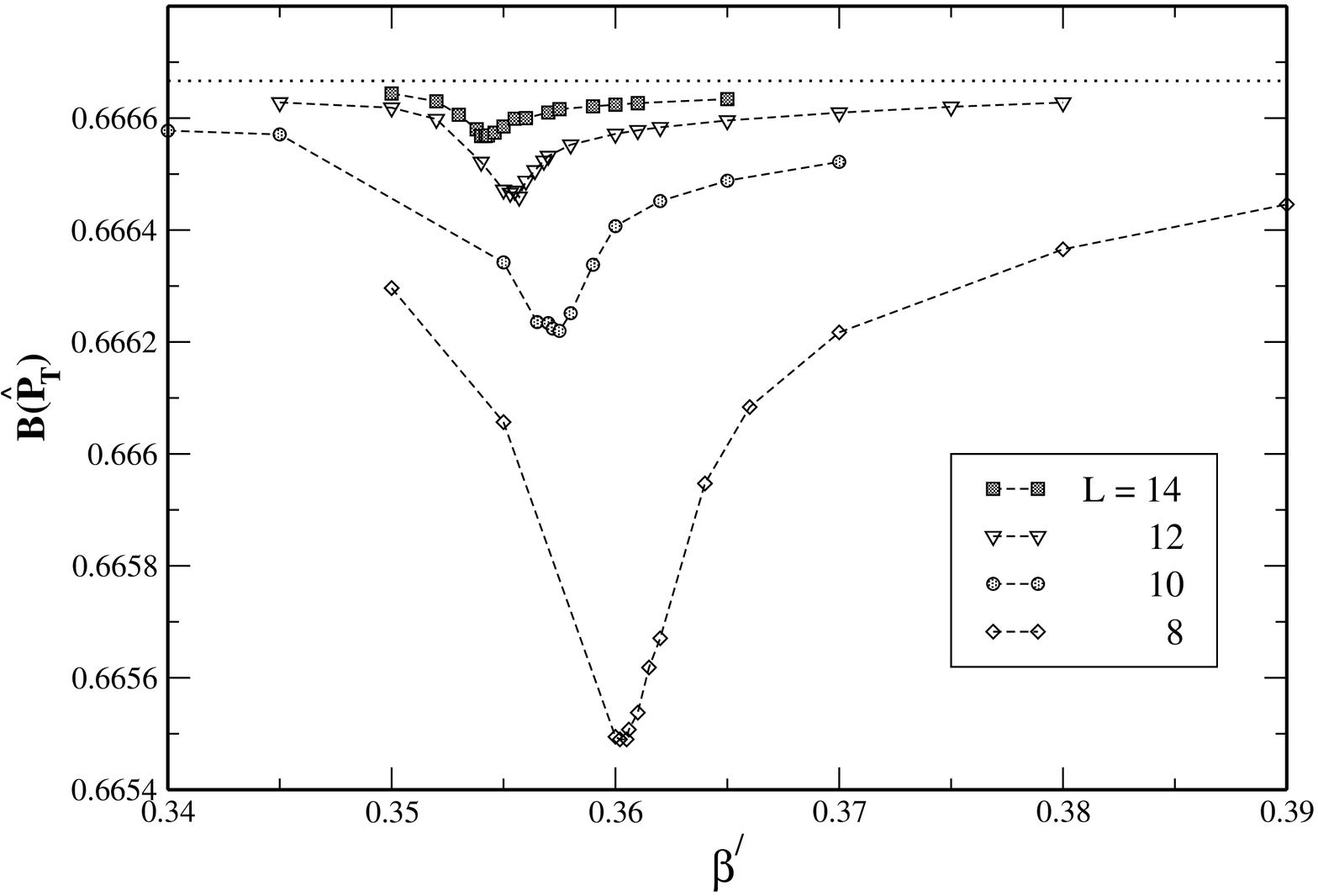}}
\caption{(a): The mean values of the transverse plaquette, $P_T$, with the volume; (b):The 
Binder cumulant, $ B(\hat{P_{T}})$,  as a function of the volume 
(the errors are included in the symbols' size and the dashed lines guide the trend).} 
\label{PT_BPT}
\end{center}
\end{figure}

We choose to keep  $\beta$ constant to the value 1.4  while we let $\beta^{'}$ vary.
In Fig.\ref{PT_BPT}(a)  we depict the transverse--like plaquette as a function of the transverse--like
coupling for four lattice volumes. One first observation is that the transition point moves to smaller 
values of $\beta^{'}$ as the volume increases. Then  there is a  difference though small  for the values  
of $P_T$  in the transition region between the two  bigger volumes. 

In Fig.\ref{PT_BPT}(b) we present our results for  the Binder cumulant, 
$ B(\hat{P_{T}})$, and for four lattice volumes. It can be 
noticed that the minimum value of the Binder cumulant for the bigger volume lies extremely close 
to the limit value $2/3$. Although this fact provides evidence for a continuous phase transition it can not 
be used as a criterion  to distinguish a second order phase transition from a crossover.

\begin{figure}[!h]
\begin{center}
\subfigure[]{\includegraphics[scale=0.30,angle=-90]{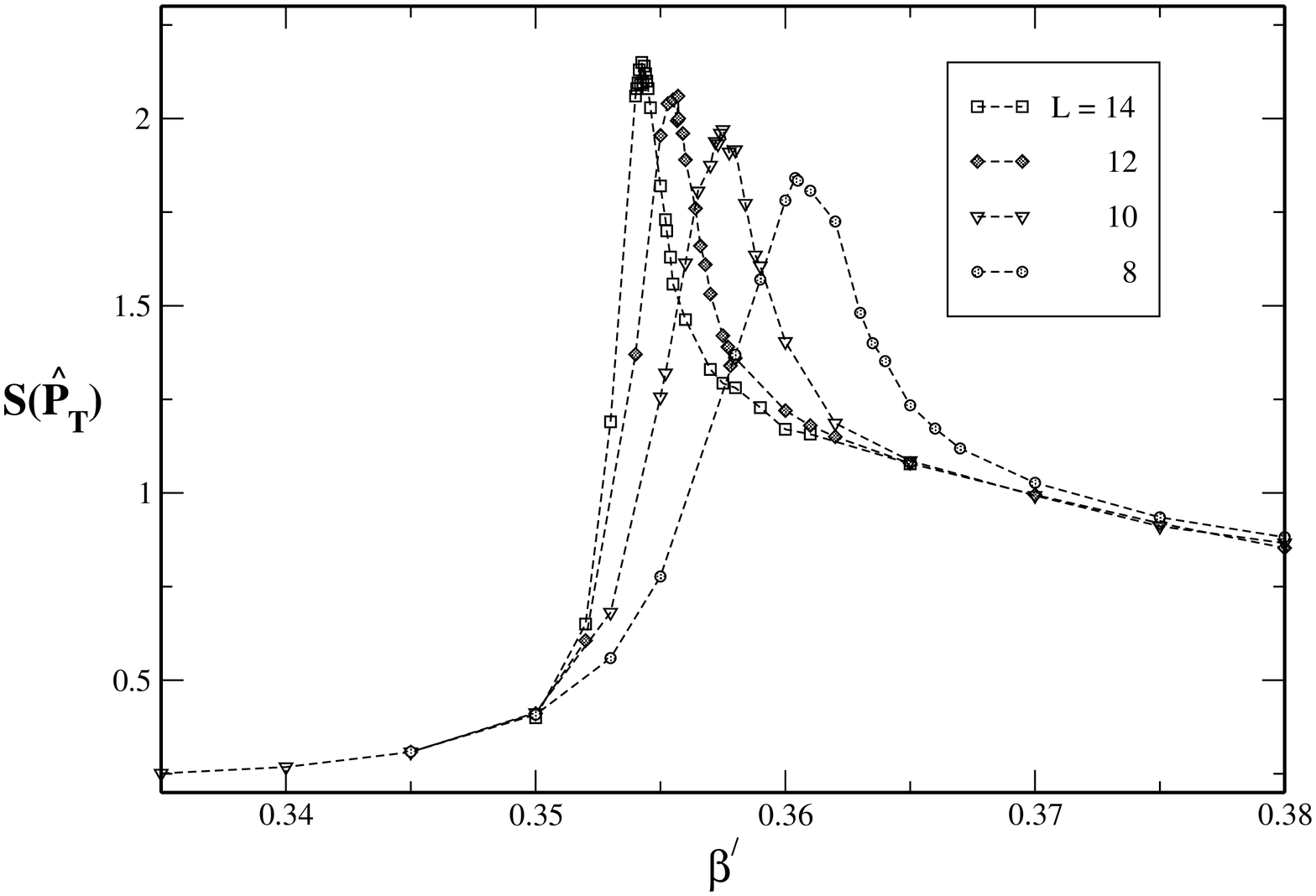}}
\subfigure[]{\includegraphics[scale=0.30,angle=-90]{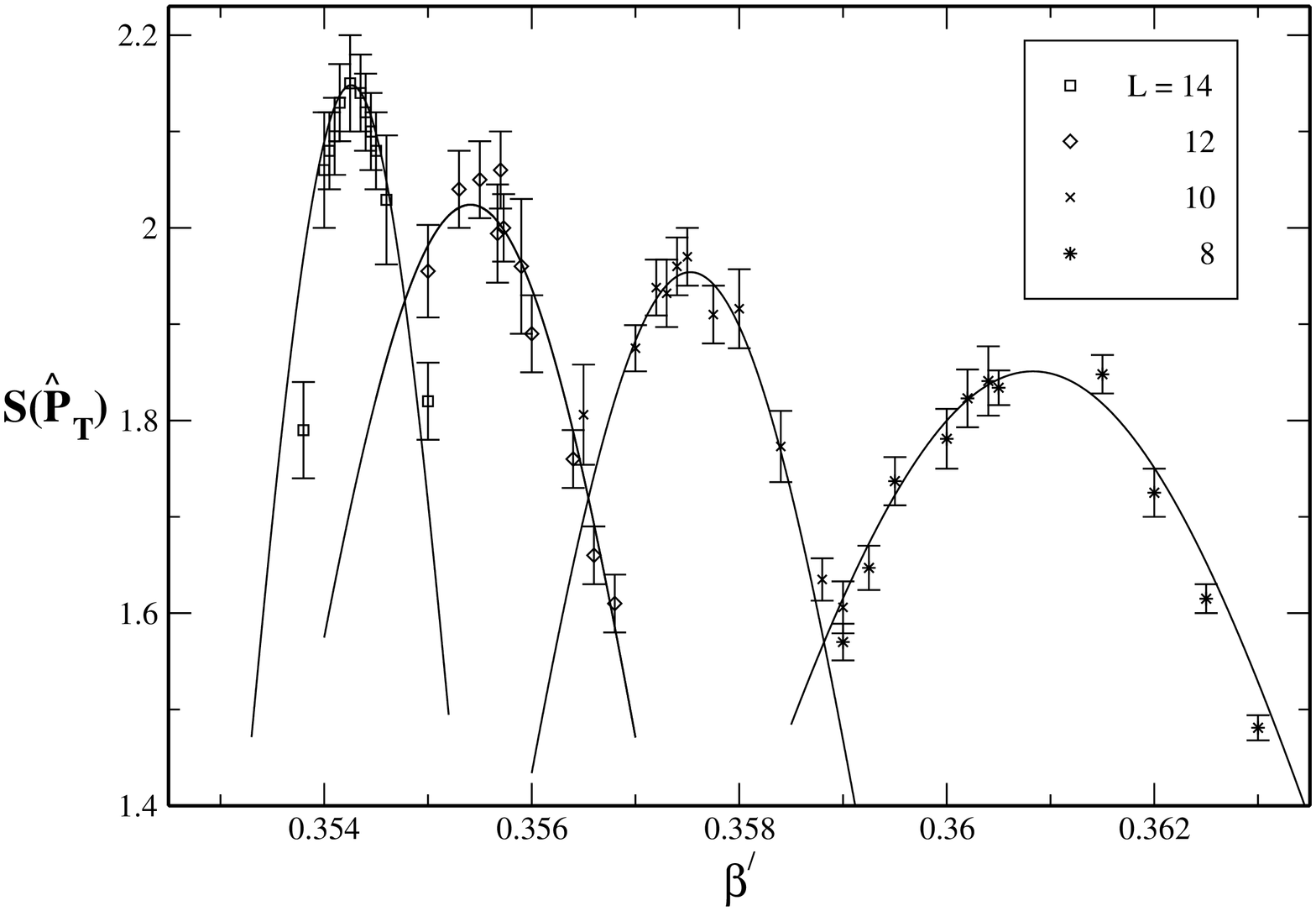}}
\caption[]{(a) The volume dependence of the susceptibility for the
transverse-like plaquette (errorbars not shown here); (b) the same with the gaussian fits shown.}
\label{PT_susc}
\end{center}
\end{figure}

The  susceptibility $S(\hat{P}_T)$ as a function of the five dimensional volume 
is depicted in Fig.\ref{PT_susc}.   

\noindent The susceptibility peaks display  a small scaling with the volume. The pseudocritical $\beta^{'}_c$ and the 
maxima of the susceptibility,  $S_{max}(\hat{P}_T)$, for each lattice volume have been estimated using 
a gaussian fit around the peak and  are given in Table \ref{beta_prime_Susc}. 
\begin{table}
\begin{center}
\begin{tabular}{ccc}
\hline \hline 
 $L$ & $\beta^{'}_c$ &  $S_{max}(\hat{P}_{T})$  \\
\hline
$8$   & 0.36083(8)  & 1.857(10)  \\
$10$  & 0.35746(6)  & 1.940(12)  \\
$12$  & 0.35541(8)  & 2.024(14)  \\
$14$  & 0.35426(3)  & 2.148(15)  \\
\hline
\end{tabular} \vspace*{0.3cm}
\caption{\label{beta_prime_Susc} The pseudocritical values $\beta^{'}_c$  
and the maxima of the transverse--like susceptibility,  $S_{max}(\hat{P}_{T})$ at $\beta=1.4$.}
\end{center}
\end{table}

\begin{figure}[!h]
\begin{center}
\includegraphics[angle=270,width=10cm]{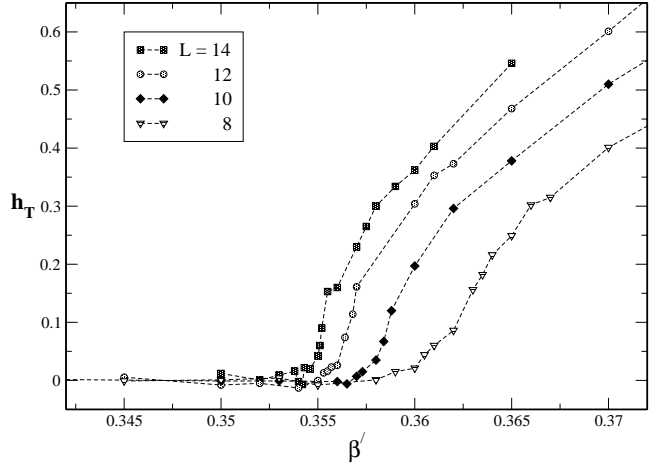}
\caption[]{The volume dependence of the transverse helicity modulus 
(the errors are  not included  and the dashed lines guide the trend).}
\label{hT}
\end{center}
\end{figure}
\newpage

The transverse helicity modulus, $h_T$, offer the advantage of rendering more clear the 
phase transition as the lattice volume increases. In the Layer phase the force 
between neighbouring layers is confining, making the system insensitive to the presence of the
external flux and thus giving a zero value to the 'transverse'
h.m. When the system passes to the Coulomb phase the force becomes
Coulomb-like and thus $h_{T}$ obtains a non-zero value. The behaviour of the space--like h.m., $h_{S}$, is 
quite different: the transition from the Layer to the Coulomb phase,  from the point of view 
of the 4--D layers, is actually a passage 
from a 4-D to a 5-D Coulomb law.  Thus it is expected that $h_S$
gets a constant value   \footnote{Indeed the value obtained by the $h_S$ is constant 
and close to one.}.
In Fig.\ref{hT} the transverse helicity modulus is shown for four lattice volumes. It can be noticed that
as the volume increases the transition to the Coulomb phase becomes steeper and allows, in principle, an  
estimation of the critical point. Indeed  
in comparison with the transverse--like plaquette (see Fig.\ref{PT_BPT}(a))  the use of the helicity 
modulus, $h_T$,  helps to get  a less unambigous signal of the phase transition. 

Now, assuming  the presence of a  second order phase transition, we  expect that near to the 
critical point  the  correlation length has to be given by the following expression:
\begin{equation}
\xi\sim |\beta^{'} - \beta^{'}_{c}|^{-\nu}
\end{equation}

\begin{figure}[!h]
\begin{center}
\subfigure[]{\includegraphics[scale=0.30,angle=-90]{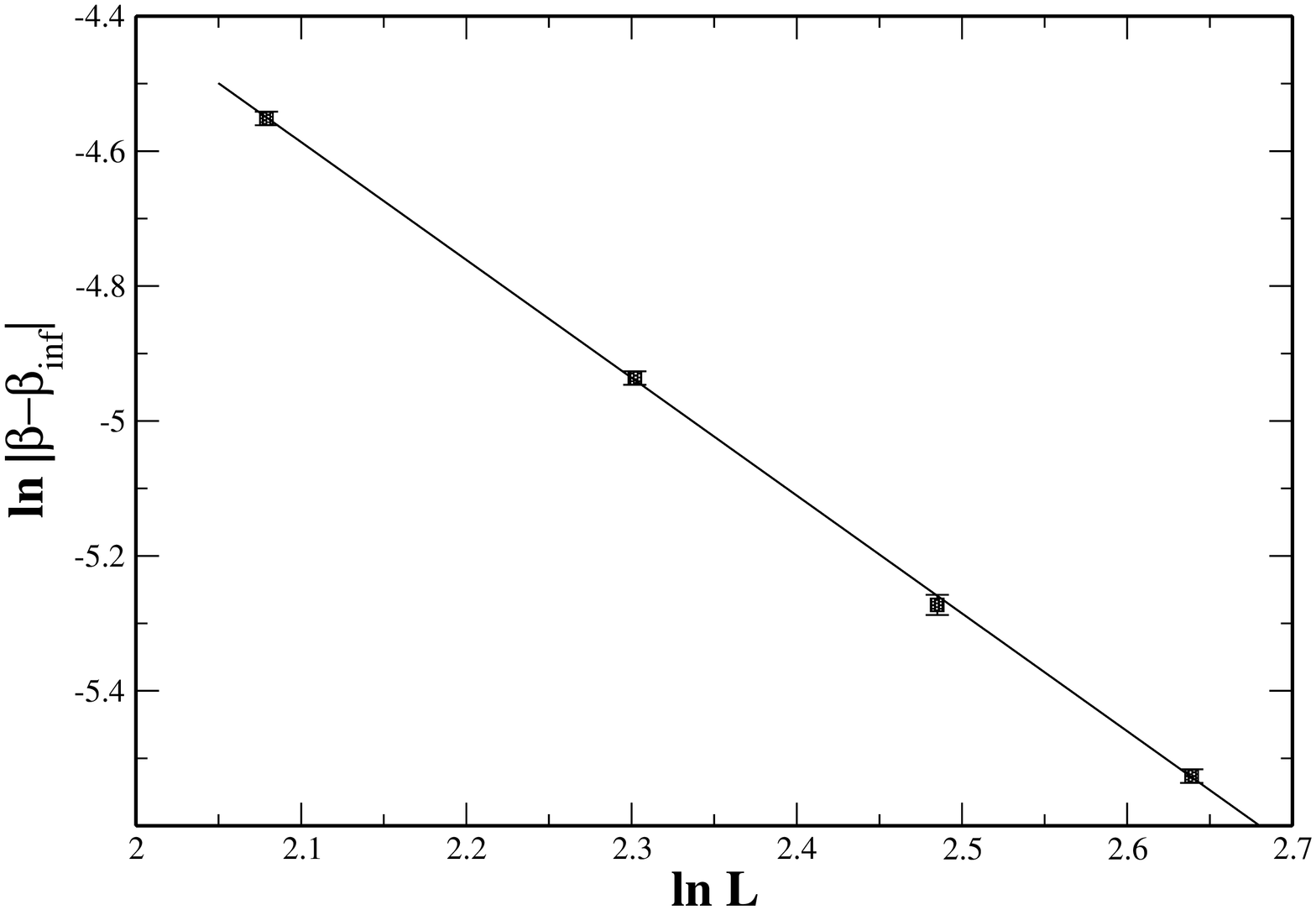}}
\subfigure[]{\includegraphics[scale=0.30,angle=-90]{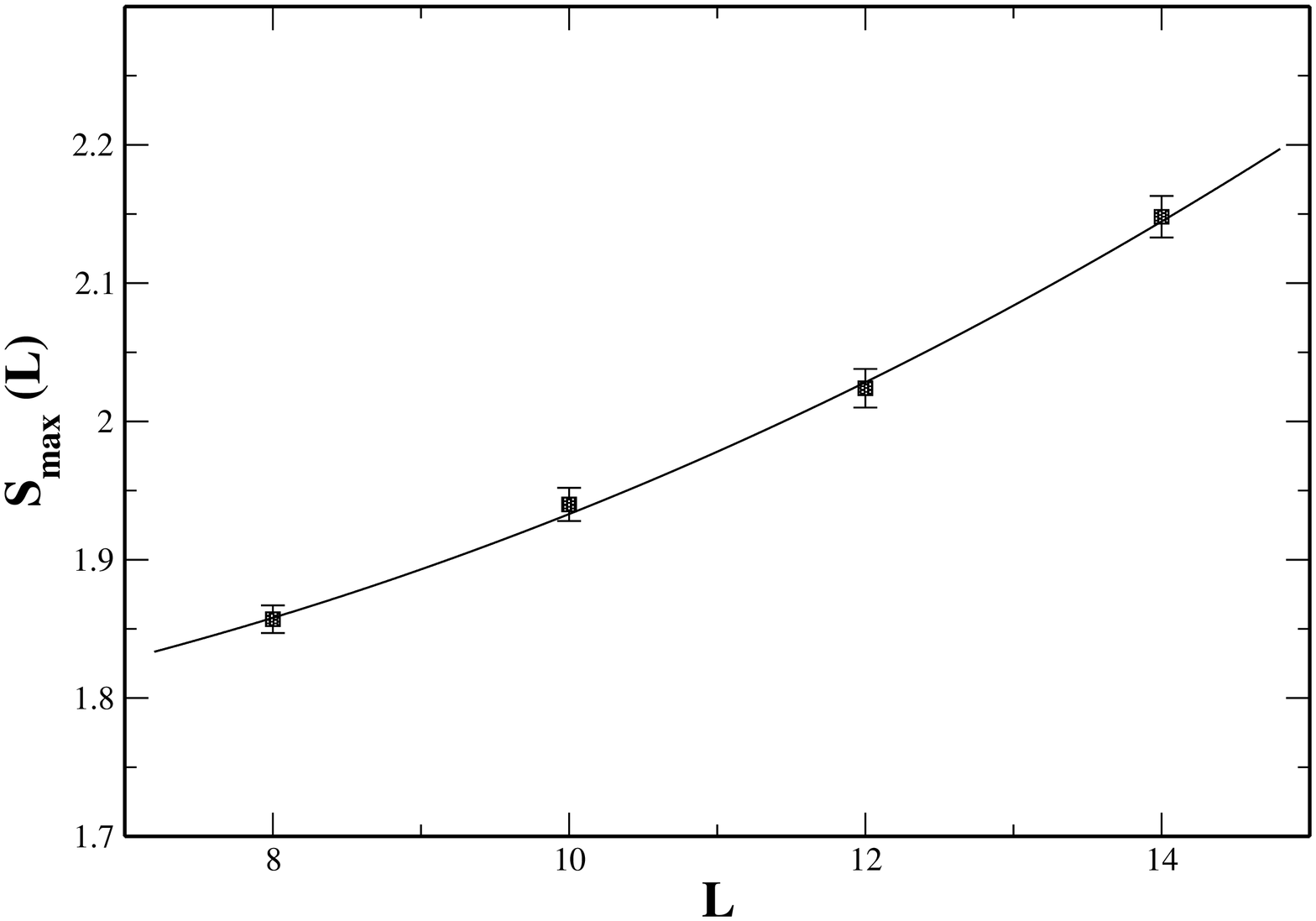}}
\caption{(a) Fit of the Eq.(\ref{beta_L}); (b) Fit of the Eq.(\ref{S_L_fit}).}
\label{beta_prime_L_susc}
\end{center}
\end{figure}

We also assume that the pseudocritical value of the transverse gauge coupling is expressed as 
a function of the lattice length according to the expression \cite{Newman}:
\begin{equation}
\beta^{'}(L)=\beta^{'}_{\infty} (1+C_{1} L^{-\frac{1}{\nu}})
\end{equation}
or equivalently by:
\begin{equation}\label{beta_L}
\ln|\beta(L)^{'}-\beta^{'}_{\infty}|=C_{2}-\frac{1}{\nu} \ln(L)
\end{equation}
Using the pseudocritical values for the gauge coupling of Table \ref{beta_prime_Susc} we obtain (see 
Fig.\ref{beta_prime_L_susc}(a)):
\begin{equation}\label{nu_beta}
\nu = 0.57(5) ~~ \mbox{and} ~~  \beta^{'}_{\infty} = 0.35028(53)
\end{equation}

The asymptotic scaling law for the susceptibility takes the form:
 \begin{equation}\label{S_L_fit}
 S_{max}(L)=C_{1}+C_{2} L^{\frac{\gamma}{\nu}}
 \end{equation}
Using  the values of Table \ref{beta_prime_Susc} we obtain $\frac{\gamma}{\nu} = 2.19(74)$ and 
with the help of the   Eq.(\ref{nu_beta}) we get  $\gamma = 1.24(44)$. In the  Fig.\ref{beta_prime_L_susc}(b)
we depict the result of the fitting procedure  using the Eq.(\ref{S_L_fit}). 
In other words we obtain a volume exponent 
equal to 0.44(15) which provides a serious evidence for a second order phase transition.

\section{Conclusions}

We  consider a  U(1) gauge model in $4+1$ dimensions with anisotropic gauge couplings. 
The main property of this model is the existence of a new phase which is 
called Layer and is characterized by Coulomb--like interaction on a 4--D subspace and confinement 
along the fifth direction. The other two phases of the phase diagram are a 5--D Coulomb phase and a 
confinement phase. The study of the phase transitions reveals that the Layer and the Confinement phases are
separated by a weak first order phase transition whose critical gauge coupling is found very close 
to the Coulomb--confinement critical coupling of  the 4--D model. 
Furthermore for the Layer--Coulomb phase 
transition we provide serious evidence of a second order phase transition. 
If this conclusion persists after the 
use of bigger lattice volumes it would provide a promising scenario for a gauge field localization 
based on a model that features  a  continuum limit. 

A final remark should be added which has to do with a possible connection of  our 5--D gauge model 
with the percolation model. In \cite{Gliozzi},\cite{Ziff}, 
it is argued that percolation in three dimensions can be viewed as a  gauge theory and it can 
capsulate most of the features of confinement and the glueball spectrum. 
The values of the exponents $\gamma$ and $\nu$ given at the end of the  Section  4.2 
are in a good 
agreement with the  values of the corresponding exponenents of the 5--D percolation model which are:
$\gamma_{5D perc}=1.18$ and $\nu_{5D perc}=0.57$ (see \cite{perc5D}). 
Although this  fact alone  can not justify any 
further argumentation on a possible universality class issues, however it might be useful
in providing a  new point of view  for the confinement  mechanism along the extra dimension.

\section{Acknowledgements}
We acknowledge support from the EPEAEK programme  ``Pythagoras II" co-funded by the European Union 
(75\%) and the Hellenic State (25\%). We are grateful to K. Anagnostopoulos 
and G. Koutsoumbas for reading and discussing the manuscript and making useful comments.

\end{document}